\documentclass[11pt,twoside,onecolumn]{article}
\usepackage[]{amsmath,amssymb}
\usepackage[]{epsfig}
\newcommand{\be}{\begin{eqnarray}}
\newcommand{\ee}{\end{eqnarray}}

\title{\bf CMB: A Look Inside the Inflaton}
\author{G.L.~Alberghi\thanks{e-mail: alberghi@bo.infn.it}$\ $
\\
{\em Dipartimento di Fisica, Universit\`a di Bologna, and}
\\
{\em Istituto Nazionale di Fisica Nucleare,
Sezione di Bologna, Italy}}
\begin{document}
%
%
\maketitle
\begin{abstract}
We show that if the field seeding the formation of  the cosmic structures is a dynamically arising bosonic condensate,
the features we observe in the CMB might be interpreted as a manifestation of its compositeness.


%
%
%
%

\end{abstract}
%
\pagestyle{plain}
\raggedbottom
\setcounter{page}{1}
%
%
%
%
%
%
%

\section{Introduction}
\label{intro}
Inflation has  become a standard ingredient for the description of the very early Universe  
~\cite{Infla}. In fact, it solves some of the problems of the standard Big-Bang
scenario and also makes predictions about cosmic microwave background radiation (CMB)
anisotropies which are being measured with higher and higher precision. 
\par
The period of accelerated expansion is usually assumed to be driven by a 
real scalar field, the Inflaton, whose quantum fluctuations become the seeds
for the formation of structures and whose signatures are encoded in the
almost scale invariant power spectrum of the CMB.
In fact Inflationary models \cite{infl} not only explain the large-scale
homogeneity and isotropy of the universe, but also provide a natural
mechanism to generate the observed magnitude of inhomogeneity.
During the period of inflation, quantum fluctuations are generated
within the Hubble horizon, and then stretch outside it to
become classical. In the subsequent deceleration phase,
these frozen fluctuations re-enter the horizon, and seed
the matter and radiation density fluctuations observed in the
universe.

With the development of cosmological observations,
such as WMAP \cite{wmapt} and Sloan Digital Sky Survey \cite{sdss}
the $\Lambda$CDM model has shown to be an excellent fit to the WMAP
three-year data. A nearly scale-invariant, adiabatic primordial
power spectrum generated during inflation can be taken as the seed
for the anisotropy of CMB. Even though a red power spectrum ($n_s < 1 $) for the
curvature perturbations is certainly a good fit to the WMAP data, a running
spectral index  with $ \alpha = d n_s / d  \log  k   $ slightly  improves
 the fit  \cite{wmapt}.
The running is allowed not only by taking into account  the WMAP data, 
but also in combination with other CMB and/or large scale structure information,
such as 2dFGRS \cite{tdfgrs} and the Sloan Digital Sky Survey
\cite{sdss}. Further analysis of a possible running spectral index
is discussed in \cite{fxy} and theoretical explorations of WMAP
results are carried out recently in \cite{hls}.
The WMAP data also implies that a red power spectrum ($n_s<1$) at
$k=0.05$Mpc$^{-1}$ is favored and the running of the spectral index
is not required at more than the $95\%$ confidence level. 
\par
The CMB data might lead one to the attempt of reconstructing to high precision
 the inflaton potential, as done in refs. \cite{Potential}.
 Further, it has been recently suggested that inflation might provide a window
towards {\em trans\/}-Planckian physics \cite{Transplanck, Branden}.
The reason for this is that inflation magnifies all quantum
fluctuations and, therefore, red-shifts originally trans-Planckian
frequencies down to the range of low energy physics.
In this framework on might think that  even non-commutativity arising 
at the highest energies (see \cite{NonComm})  might play a role in determining the CMB features.
 \par
 In this article we would like to suggest a further possible interpretation of the CMB features,
 by  assuming that the inflaton might not be a fundamental field but a dynamically
 arising bosonic condensate, or at least that this condensate 
 could  be at the origin of  the CMB anisotropies.
 In refs. \cite{Giacosa:2008rw, Alexander:2008vt}  it is actually shown that 
  a four-fermion interaction arises in General Relativity when fermions are covariantly coupled, and
 that at early times and high densities the correlation between pairs of fermions is enhanced. 
This enhancement leads to a BCS-like condensation of the fermions and opens a gap,
indicating the dynamical formation of a bosonic condensate.
If this is the case, it is possible that the features of the present day CMB are a sign
of the compositeness of the inflaton (or curvaton) field. 
The correlation functions of a scalar field usually employed in the derivation the CMB power spectrum
should then be substituted by the correlation functions for a composite
scalar field operator in the form, for example, $ \bar \psi \psi $ or $ i \bar \psi \gamma_5 \psi $.
The leading behaviour of these correlation functions would be, as usual,
those of a fundamental scalar field in a de Sitter spacetime,
leading to a scale invariant spectrum. 
 On the other hand the deviations from a scale invariant power spectrum, described by the spectral index 
 $n_s$ and its running $\alpha$, might be the sign of the compositeness of the seeding field.
 By assuming a logarithmic correction to the two-point function (this is our ansatz, 
 derived from the literature on composite scalar field correlation functions in
 flat spacetime \cite{Compositeness}) we were able to derive values for
 spectral index and  its running compatible with  the experimental data.
 
  \section{Cosmological Fermionic Condensate}

\label{subsec:4fermi}
In this subsection  we outline the essential steps taken in \cite{Giacosa:2008rw} to describe the
cosmological condensation.
Let us start by considering purely gravitational 
dynamics as given by the Holst action \cite{Holst1996}
\begin{equation}\label{faction}
   S[e,A] = \frac{1}{16\pi G} \left( \int d^4x\,
   e\,e_a^\mu\,e_b^\nu\,F_{\mu\nu}^{ab} 
   - \frac{1}{\gamma} \int d^4x\,e\,e_a^\mu\,e_b^\nu\, 
   \tilde{F}_{\mu\nu}^{ab} \right) ,
\end{equation}
which is a functional of the tetrad field $e_\mu^a$ \cite{Ashtekar1991}. 
Here $a=0,1,2,3$ is the internal Lorentz index, 
$\mu=0,1,2,3$ the coordinate index, and $e\equiv\det e^\mu_a$. $F_{\mu\nu}^{ab}$
is the curvature of the connection $A_{cd}^\mu$ defined as 
\begin{equation}\label{conn}
   A_{cd}^\mu 
   = e_c^\nu \left( e_{d,\nu}^\mu - \Gamma_{\rho\nu }^\mu\,e_d^\rho
   \right) ,
\end{equation}
and $\Gamma_{\rho\nu}^\mu$ are the Christoffel symbols and finally
$\tilde{F}_{\mu\nu}^{ab}=\frac12\epsilon^{ab}_{\ \ cd}\,F_{\mu\nu}^{cd}$ is the (internal) dual field strength. 


The first term in (\ref{faction}) yields the tetrad formulation
(Palatini action) of the Einstein-Hilbert action, the latter emerging when inserting a solution to the 
associated equation of motion ($A_{cd}^\mu$ being a torsion-free spin connection
 $\omega_{cd}^\mu[e]$) into (\ref{faction}) and using $g_{\mu\nu}=e_\mu^a\,e_{\nu a}$.
The second term is identically zero due to the Bianchi identity for the Riemann tensor.
It follows that, regardless of the value of $\gamma$, the action (\ref{faction}) is \emph{classically\/} 
equivalent to the familiar Einstein-Hilbert action \cite{Perez2006}. 
What holds true for pure gravity is no longer valid if minimally coupled chiral 
fermions are introduced. The equation of motion for the
tetrad $e_a^\mu$ subject to a fermionic source is solved in terms of a
connection $A_{cd}^\mu$ having two contributions, a torsion-free spin
connection for $e_a^\mu$ (as in the purely gravitational case) and a
torsion term related to the axial fermion current. Upon substituting $A_{cd}^\mu$ back into the action, 
a four-fermion interaction of the following form emerges \cite{Perez2006}: 
\begin{equation}\label{fermint}
   S_{\mathrm{int}} = \frac{K}{2} \int d^4x\,e \left(
    \bar\psi\gamma_5\gamma_a\psi \right) 
    \left( \bar\psi\gamma_5\gamma^a\psi \right) \,; \qquad 
   K = - 3\pi G\,\frac{\gamma^2}{\gamma^2+1} 
    = - \frac{9}{8M_P^2}\,\frac{\gamma^2}{\gamma^2+1} \,.
\end{equation}
Thus the Immirzi parameter acquires physical relevance through the presence of massless fermions, 
even though gravity is still treated classically.


The action describing the fermions reads 
\begin{equation}
   S_{\mathrm{ferm}} 
   = \int d^4x\,e \left[ \bar\psi\,ie_a^\mu\gamma^a D_\mu[e]\,\psi 
   + \frac{K}{2} \left( \bar\psi\gamma_5\gamma_a\psi \right) 
   \left( \bar\psi\gamma_5\gamma^a\psi \right) \right] ,
\end{equation}
where $D_\mu[e]$ is the covariant derivative with respect to the connection $A$. 
Let us consider the system of interacting
 fermions in a de Sitter spacetime in FRW coordinates 
\begin{equation}  \label{metric}
ds^2 = dt^2 - a^2(t)\,d\vec{x}\cdot d\vec{x} \,,
\end{equation}
where $a=a_{0}\,e^{Ht}$ is the scale factor. In this case the vierbein reads 
\begin{equation}\label{tetrad}
   e_{\mu a} = \delta_{\mu 0}\,\delta_{a0} - a(t)\,\delta_{\mu i}\,
   \delta_{a i} \,.
\end{equation}
The consideration of a de Sitter spacetime is justified in an epoch
where the energy density belonging to fluctuating degrees of freedom is sufficiently diluted as compared to
 the energy density of condensed degrees of freedom.

Applying a Fierz transformation to the current-current interaction in (\ref{fermint}) yields
\begin{equation}\label{fierz}
   \left( \bar\psi\gamma_5\gamma_a\psi \right) 
   \left( \bar\psi\gamma_5\gamma^a\psi \right) 
   \to \frac{1}{N} \left( \bar\psi\psi \right)^2 
   + \frac{1}{N} \left( \bar\psi i\gamma_5\psi \right)^2 + \dots \,,
\end{equation}
where the dots refer to flavor-nonsinglet contributions and products of vector and axial-vector currents, which do not lead to vacuum condensates. Allowing for a bare cosmological constant $\Lambda_0$, and denoting the bare reduced Planck mass by $M_0$, the complete action then takes the form
\begin{equation}\label{Stot}
   S = \int d^4x\,e \left\{ M_0^2 H^2  - \Lambda_0
   + \bar\psi\,ie_a^\mu\gamma^a D_\mu[e]\,\psi + \frac{K}{2N} 
   \left[ \left( \bar\psi\psi \right)^2 
   + \left( \bar\psi i\gamma_5\psi \right)^2 + \dots \right] 
   \right\} ,
\end{equation}
where only the scalar flavor-singlet bilinears are shown explicitly. 
For condensation to take place $K$ needs to be positive, implying an imaginary 
$\gamma$ with $|\gamma|<1$. In fact, only if $K>0$ an attractive force occurs 
between fermions in the scalar channel \cite{njl}. In this case bound states form: 
The operator $\bar\psi\psi$ corresponds to a scalar field $\sigma$, 
while $\bar\psi i\gamma_5\psi$ corresponds to a pseudoscalar field $\pi$. 
In addition, $(N^2-1)$ flavor-nonsinglet scalar and pseudoscalar fields 
$\sigma^k\sim(\bar\psi t^k\psi)$ and $\pi^k\sim(\bar\psi t^k i\gamma_5\psi)$ 
appear. In the usual treatment
of the Nambu--Jona-Lasinio model in Minkowski space \cite{njl} the dynamical
 breaking of chiral symmetry occurs for sufficiently large $K$. 
 As a result, the scalar fields $\sigma$ and $\sigma^k $ become massive, while the 
 pseudoscalar fields $\pi$ and $\pi^k$ remain massless and represent 
 Goldstone bosons (see, e.g., \cite{Klevansky} for a review).

In a de Sitter background dynamical chiral symmetry breaking occurs for all values $K>0.$ 
Only the isosinglet fields are considered here. In fact, only the isosinglet scalar 
field $\sigma$ can acquire a nonzero vacuum expectation value and thus is relevant 
for de Sitter cosmology in the early Universe. In a de Sitter 
spacetime we expect that the particles associated with the remaining low-lying flavor-nonsinglet
fields are sufficiently diluted to provide for the self-consistency of the de Sitter geometry.

\section{Two Point Function in de Sitter Space-Time}

%


Let us consider a scalar field $ \phi $. We
usually assume that it can be split into a classical
background piece and a piece due to fluctuations according to
\be
\phi  \left(  \tau,\mathbf{x}\right)   = \phi^{\left(  0\right)  }  \left(  \tau \right)+\delta\phi \left(  \tau,\mathbf{x}\right)  .
\ee 
From now on we forget about the homogeneous part and  assume that scalar field fluctuations 
are at the origin of the CMB we observe today.
For convenience we have introduced the conformal
time $ \tau $, such that the metric is given by
\begin{equation}
ds^{2}=a\left(  \tau\right)  ^{2}\left(  dt^{2}-d\mathbf{x}^{2}\right)  .
\end{equation}
In these coordinates the Klein-Gordon equation, ignoring the potential
piece, becomes
\begin{equation}
\delta\phi_{\mathbf{k}}^{\prime\prime}+2\frac{a^{\prime}}{a}\delta
\phi_{\mathbf{k}}^{\prime}+k^{2}\delta\phi_{\mathbf{k}}=0,
\end{equation}
where we have Fourier transformed in space and introduced the comoving
momentum $\mathbf{k}$. The conventions are such that%
\begin{equation}
\delta\phi\left(  \mathbf{x}\right)  =\frac{1}{\left(  2\pi\right)  ^{3/2}%
}\int\delta\phi_{\mathbf{k}}e^{i\mathbf{k}\cdot\mathbf{x}}d^{3}k.
\end{equation}
We have also introduced the notation $\prime$ for derivatives with respect to
conformal time. If we then introduce the rescaled field $\mu=a\delta\phi$, the
equation becomes
\begin{equation}
\mu_{\mathbf{k}}^{\prime\prime}+\left(  k^{2}-\frac{a^{\prime\prime}}%
{a}\right)  \mu_{\mathbf{k}}=0. \label{modeeq}%
\end{equation}
To proceed, we assume that the scale factor depend on conformal time as%
\begin{equation}
a\sim\tau^{1/2-\nu}, \label{aeta}%
\end{equation}
where $\nu$ is a constant. An important example is $a\sim e^{Ht}$ with
$H=\mathrm{const.}$, where the change of coordinates gives%
\begin{equation}
\frac{d\tau}{dt}=\frac{1}{a\left(  t\right)  }=e^{-Ht}\Longrightarrow a\left(
\tau\right)  =-\frac{1}{H\tau},
\end{equation}
and we find that $\nu=\frac{3}{2}$. Note that the physical range of $\tau$ is
$-\infty<\tau<0$. The equation for the fluctuations, with $a$ of the form
above, becomes%
\begin{equation}
\mu_{\mathbf{k}}^{\prime\prime}+\left(  k^{2}-\frac{1}{\tau^{2}}\left(
\nu^{2}-\frac{1}{4}\right)  \right)  \mu_{\mathbf{k}}=0.
\end{equation}
This is a well known equation which is solved by Hankel functions.
The general solution is given by%
\begin{equation}
f_{k}\left(  \tau\right)  =\frac{\sqrt{-\tau\pi}}{2}\left(  C_{1}\left(
k\right)  H_{v}^{\left(  1\right)  }\left(  -k\tau\right)  +C_{2}\left(
k\right)  H_{v}^{\left(  2\right)  }\left(  -k\tau\right)  \right)  ,
\label{hank}%
\end{equation}
where $C_{1}\left(  k\right)  $ and $C_{2}\left(  k\right)  $ are to be
determined by initial conditions.

When quantizing this system one needs to introduce oscillators $a_{k}\left(
\tau\right)  $ and $a_{-k}^{\dagger}\left(  \tau\right)  $ such that%

\begin{align}
\mu_{\mathbf{k}}\left(  \tau\right)   &  =\frac{1}{\sqrt{2k}}\left(
a_{k}\left(  \tau\right)  +a_{-\mathbf{k}}^{\dagger}\left(  \tau\right)
\right) \label{mupi}\\
\pi_{\mathbf{k}}\left(  \tau\right)   &  =\mu_{\mathbf{k}}^{\prime}\left(
\tau\right)  +\frac{1}{\tau}\mu_{\mathbf{k}}\left(  \tau\right)
=-i\sqrt{\frac{k}{2}}\left(  a_{k}\left(  \tau\right)  -a_{-\mathbf{k}%
}^{\dagger}\left(  \tau\right)  \right)  ,\nonumber
\end{align}
obey standard commutation relations. These
oscillators are time dependent, and can be expressed in terms of oscillators
at a specific moment in time using the Bogolubov transformations
\begin{align}
a_{\mathbf{k}}\left(  \tau\right)   &  =u_{k}\left(  \tau\right)
a_{\mathbf{k}}\left(  \tau_{0}\right)  +v_{k}\left(  \tau\right)
a_{-\mathbf{k}}^{\dagger}\left(  \tau_{0}\right) \label{oscutv}\\
a_{-\mathbf{k}}^{\dagger}\left(  \tau\right)   &  =u_{k}^{\ast}\left(
\tau\right)  a_{-\mathbf{k}}^{\dagger}\left(  \tau_{0}\right)  +v_{k}^{\ast
}\left(  \tau\right)  a_{\mathbf{k}}\left(  \tau_{0}\right)  ,\nonumber
\end{align}
where
\begin{equation}
\left\vert u_{k}\left(  \tau\right)  \right\vert ^{2}-\left\vert v_{k}\left(
\tau\right)  \right\vert ^{2}=1.
\end{equation}
The latter equation makes sure that the canonical commutation relations are
obeyed at all times if they are obeyed at $\tau_{0}$. We can now write down
the quantum field%
\begin{equation}
\mu_{\mathbf{k}}\left(  \tau\right)  =f_{k}\left(  \tau\right)  a_{\mathbf{k}%
}\left(  \tau_{0}\right)  +f_{k}^{\ast}\left(  \tau\right)  a_{-\mathbf{k}%
}\left(  \tau_{0}\right)  ,
\end{equation}
where%
\begin{equation}
f_{k}\left(  \tau\right)  =\frac{1}{\sqrt{2k}}\left(  u_{k}\left(
\tau\right)  +v_{k}^{\ast}\left(  \tau\right)  \right)
\end{equation}
is given by (\ref{hank}).

The usual procedure for fixing for the initial conditions is to consider the
infinite past and choose a state annihilated by the annihilation operator,
i.e.%
\begin{equation}
a_{\mathbf{k}}\left(  \tau_{0}\right)  \left\vert 0,\tau_{0}\right\rangle =0,
\label{vakekv}%
\end{equation}
for $\tau_{0}\rightarrow-\infty$.  From (\ref{mupi}) we conclude that
\begin{equation}
\pi_{\mathbf{k}}\left(  \tau_{0}\right)  =-ik\mu_{\mathbf{k}}\left(  \tau
_{0}\right)  ,
\end{equation}
for $\tau_{0}\rightarrow-\infty$. Since the Hankel functions asymptotically
behave as%
\begin{align}
H_{v}^{\left(  1\right)  }\left(  -k\tau\right)   &  \sim\sqrt{\frac{2}%
{-k\tau\pi}}e^{-ik\tau}\nonumber\\
H_{v}^{\left(  2\right)  }\left(  -k\tau\right)   &  \sim H_{v}^{\left(
1\right)  \ast}\left(  -k\tau\right)  ,
\end{align}
we find that the vacuum choice correspond to the choice $C_{2}\left(
k\right)  =0$ (and $\left\vert C_{1}\left(  k\right)  \right\vert =1$).

We have now fully determined the quantum fluctuations, and it is time to
deduce what their effect will be on the CMBR. To do this, we compute the size
of the fluctuations according to
\be
\langle \hat \varphi (\eta, {\bf x}) \hat \varphi (\eta, {\bf x'})
\rangle = \int   \frac{dk}{k} 
\frac{\sin k|{\bf x} - {\bf x'}|}{k \, |{\bf x} - {\bf x'}|}
\frac{k^3 | \varphi_k (\eta)|^2}{2 \pi^2}
\equiv \int   \frac{dk}{k} 
\frac{\sin k|{\bf x} - {\bf x'}|}{k \, |{\bf x} - {\bf x'}|}
P_\varphi (k,\eta) \,.
\label{twopoint}
\ee
from which
\begin{equation}
P\left(  k\right)  =\frac{4\pi k^{3}}{\left(  2\pi\right)  ^{3}}\left\langle
\left\vert \delta\phi_{\mathbf{k}}\right\vert ^{2}\right\rangle =\frac{k^{3}%
}{2\pi^{2}}\frac{1}{a^{2}}\left\langle \left\vert \mu_{\mathbf{k}}\right\vert
^{2}\right\rangle =\frac{k^{3}}{2\pi^{2}}\frac{1}{a^{2}}\left\vert
f_{k}\right\vert ^{2}=\frac{k^{3}}{2\pi^{2}}\frac{1}{a^{2}}\frac{\left\vert
-\tau\pi\right\vert }{4}\left\vert H_{v}^{\left(  1\right)  }\left(
-k\tau\right)  \right\vert ^{2}.
\end{equation}
This we should evaluate at late times, that is, when $\tau\rightarrow0$. In
this limit the Hankel function behaves as%
\begin{equation}
H_{v}^{\left(  1\right)  }\left(  -k\tau\right)  \sim\sqrt{\frac{2}{\pi}%
}\left(  -k\tau\right)  ^{-\nu},
\end{equation}
and we find%
\begin{equation}
P\sim\frac{1}{4\pi^{2}}\frac{1}{a^{2}}\left(  -\tau\right)  ^{1-2\nu}%
k^{3-2\nu}\sim\frac{1}{4\pi^{2}}H^{2}k^{3-2\nu}.
\end{equation}
Here we have used (\ref{aeta}) to get rid off the $\tau$ dependence.
Furthermore, if $\nu\sim3/2$ and we have a slow roll, $\ H$ is nearly constant
and can be used to set the scale of the fluctuations. In particular, we find
the well known scale invariant spectrum if $\nu=3/2$,%
\begin{equation}
P=\frac{1}{4\pi^{2}}H^{2}. \label{pt}%
\end{equation}

%
%

\section{The Condensate Corrections}

Let us  assume that the previously described condensation mechanism is realized in the
inflationary expansion of the Universe and that
the dynamically generated composite scalar field  is at the origin of the curvature perturbations.
The composite nature of this field would then become apparent in its correlation functions.
In particular the equal time 2 point correlation function discussed above 
$\langle \hat \varphi (\eta, {\bf x}) \hat \varphi (\eta, {\bf x'})
\rangle$ would be replaced by the 2-point function for $ \bar \psi \psi$
so that the power spectrum will be defined by
\be
\langle (\bar \psi \psi) (\eta, {\bf x}) (\bar \psi \psi)(\eta, {\bf x'})
\rangle 
\equiv \int \frac{dk}{k} 
\frac{\sin k|{\bf x} - {\bf x'}|}{k \, |{\bf x} - {\bf x'}|}
P (k,\eta) \,.
\label{twopointcond}
\ee
This would lead to corrections to the pure scalar field correlation function, as we expect
the leading behavior not to be affected.
We thus assume that the Fourier transform of the 2-point function will  be modified as,
factorizing the leading behavior,
\be
   \tilde \xi (k) \simeq {1 \over k^3} \left(  1 + \delta \tilde \xi \right)    
\ee
Given the relation
\be
   P(k) = {k^3 \over 2 \pi} \tilde \xi (k)  \simeq k^{n_s -1}
\ee
one obtains 
\be
    \log  \left(  1 + \delta \tilde \xi \right)  \simeq \left( n_s -1 \right) \log k
\ee
By referring to the literature (see \cite{Compositeness}) we make the ansatz for the corrections to $ \tilde \xi $ 
\be
      \delta \tilde \xi =   \log  \left[   1 + A \log \left(   {B^2 \over k^2} \right)\right]
\ee 
where  $ A $ is the amplitude of the logarithmic correction
and $ B $ is the unit  in which $ k $ is measured. This leads to an expression for the spectral index
\be
   n_s  \simeq  1 + {  \log  \left[ 1 + A \log  \left(  {B^2 \over k^2 }\right) \right] \over \log \left( {k \over B}\right)}
\ee
The behavior of the spectral index is shown in Fig. 1 for  positive A and in Fig. 2 for negative A.
In the first case the spectral index is greater than 1 (blue spectrum) whereas in the second case it is 
lower than 1 (red spectrum). In both cases the running of the spectral index
$ \alpha $ can be as large  $ \alpha \simeq -0.1 $ and is always negative.
If one would like to accomodate a blue spectrum for small $ k $ 
and a red spectrum for high $ k $ one would be required to allow for a dependence (running)  of  A on $ k $
or allow for a more generic dependence of the 2-point function correction.



%

\begin{figure}
\label{grafico1}
\centerline{
\epsfxsize=240pt \epsfbox{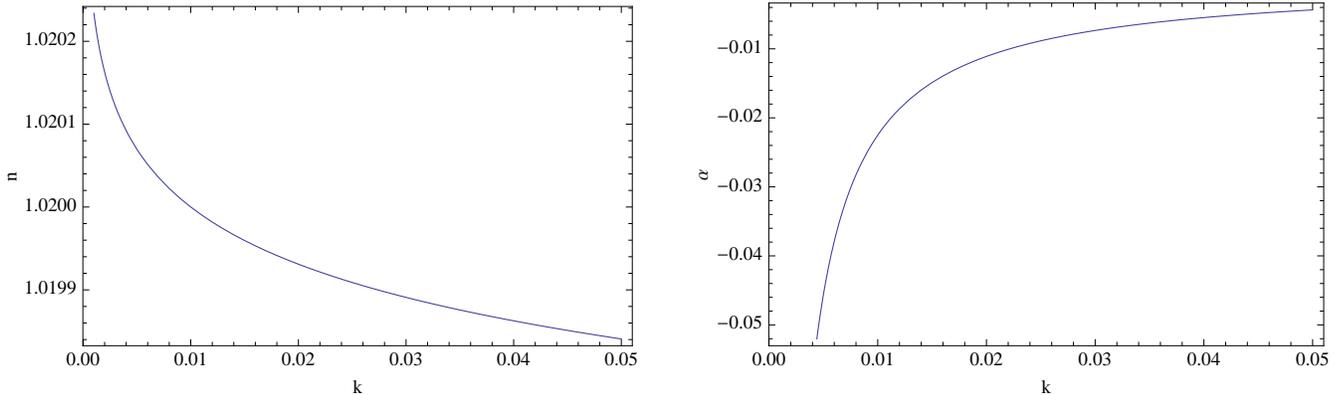}
\hspace{0.5cm}
\epsfxsize=240pt \epsfbox{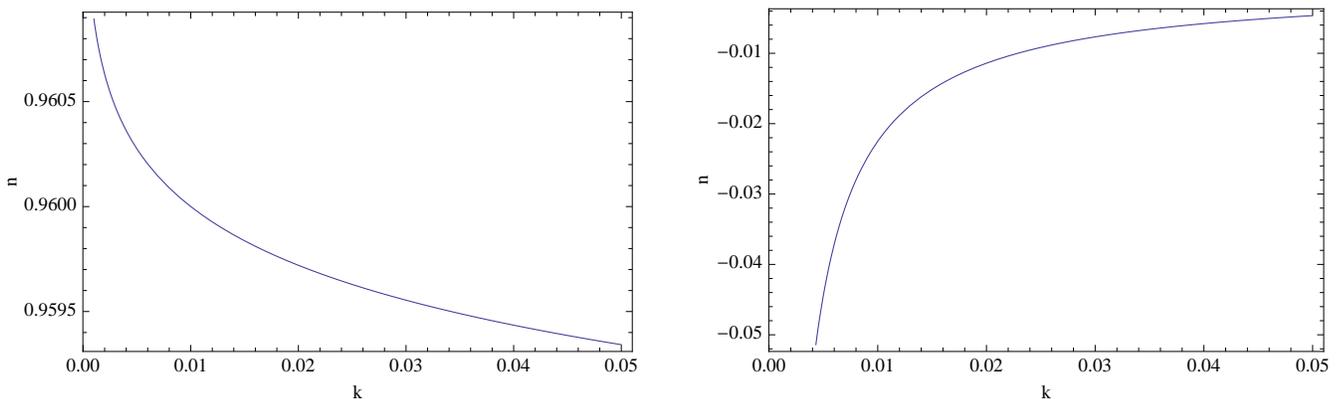}
}
\par
\noindent
%
\caption{The spectral index $n_s$ and its running $\alpha$ for $ A = -0.01$ and $ B = 0.01$  }
\end{figure}

\begin{figure}
\label{grafico2}
\centerline{
\epsfxsize=240pt \epsfbox{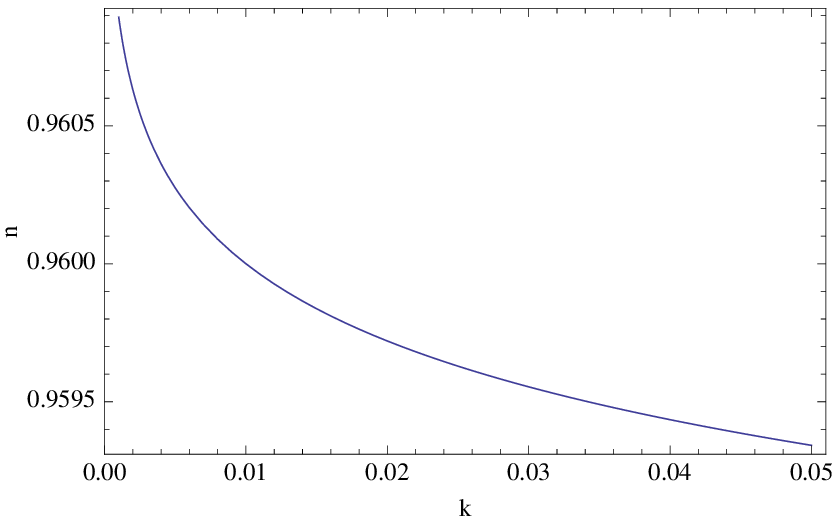}
\hspace{0.5cm}
\epsfxsize=240pt \epsfbox{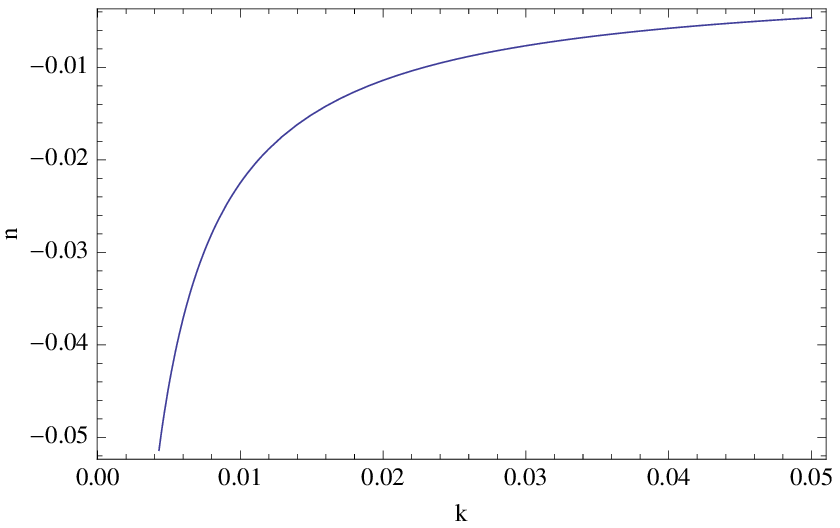}
}
\par
\noindent
%
\caption{The spectral index $n_s$ and its running $\alpha$ for $ A = 0.01$ and $ B = 0.01$ }
\end{figure}

\section{Conclusions}
We have shown that by assuming the curvature perturbations at the origin of the
CMB are due to a dynamically arising bosonic condensate, the features observed in the power spectrum
can be interpreted as an expression of its composite nature.
In particulare we were able to derive a red power spectrum compatible with the observed CMB data
with a negative running which can be of the order of magnitude $ \alpha \simeq - 0.05 $.

\label{conc}

\par

\end{document}